\documentclass[aps,pre,twocolumn,amssymb,superscriptaddress]{revtex4-2}
\usepackage{float}
\usepackage{graphicx}
\usepackage{epsfig}
\usepackage{epstopdf}
\usepackage{subfigure}
\usepackage{enumerate}
\usepackage{physics}
\usepackage{amssymb}
\usepackage{color}
\usepackage{rotating}
\begin{document}

\title{The origin of loose bound of the thermodynamic uncertainty relation in a dissipative two-level quantum system}
\author{Davinder Singh}
\affiliation{Korea Institute for Advanced Study, Seoul 02455, Korea}
\author{Changbong Hyeon}
\thanks{hyeoncb@kias.re.kr}
\affiliation{Korea Institute for Advanced Study, Seoul 02455, Korea}




\begin{abstract}
The thermodynamic uncertainty relations (TURs), originally discovered for classical systems, dictate the trade-off between dissipation and fluctuations of irreversible current, specifying a minimal bound that constrains the two quantities. In a series of effort to extend the relation to the one under more generalized conditions, it has been noticed that the bound is less tight in open quantum processes. To study the origin of the loose bounds, we consider an external field-driven transition dynamics of two-level quantum system weakly coupled to the bosonic bath as a model of open quantum system. The model makes it explicit that the imaginary part of quantum coherence, which contributes to dissipation to the environment, is responsible for loosening the TUR  bound by suppressing the relative fluctuations in the irreversible current of transitions, whereas the real part of the coherence tightens it. Our study offers a better understanding of how quantum nature affects the TUR bound.  
\end{abstract}
\maketitle

\section{Introduction}
Thermodynamic uncertainty relations (TURs) offer quantitative ideas of how much the irreversible current of a certain dynamical process and its fluctuations constrain the entropy production, which makes the lower bound of total entropy production dictated by the second law of thermodynamics more precise \cite{barato2015PRL,gingrich_dissipation_2016,dechant2018PRE,horowitz_thermodynamic_2020}.
The relation was originally derived for classical systems, particularly for continuous time Markov jump processes on networks and for overdamped Langevin dynamics in nonequilibrium steady states (NESS)  \cite{barato2015PRL,gingrich_dissipation_2016,horowitz_thermodynamic_2020}, and it has been extended to the one at finite-time \cite{pietzonka2017PRE,horowitz2017PRE,dechant2018JSM} and discrete-time Markov processes \cite{Proesmans:2017}.  
The significance of TURs has been illuminated in specific contexts of biological processes \cite{pietzonka2016JSM,hwang_energetic_2018,marsland2019JRSI,Song2020JPCL,Mugnai2020RMP,kim2021JPCB,Song2021JCP}, heat engines \cite{dechant_multidimensional_2018,pietzonka_universal_2018,holubec_cycling_2018,dechant_current_2018}, and other dynamical processes  \cite{pigolotti_generic_2017,hyeon_physical_2017,brandner_thermodynamic_2018,lee_thermodynamic_2018}. 
More recently, the universal bound of TUR has been used to infer the dissipation rate from the fluctuating currents of dynamical processes \cite{li2019NatComm}. 
The originally proposed relation has been extended to broader range of nonequilibrium processes \cite{macieszczak_unified_2018,timpanaro_thermodynamic_2019,pietzonka2017PRE,horowitz2017PRE,chiuchiu_mapping_2018,koyuk2018JPA,barato_bounds_2018,hasegawa2019PRL}, including those generated in underdamped conditions \cite{fischer2020PRE}, under periodic drives \cite{Proesmans:2017,koyuk2019PRL}, and driven by velocity-dependent forces \cite{chun2019PRE,lee2019PRE}. 
However, these processes are featured with the bounds less tight than that of the original TUR. 

The TUR has been explored for quantum systems as well \cite{carollo2019PRL,hasegawa2021PRL,ptaszynski_coherence-enhanced_2018,agarwalla_assessing_2018,liu2019thermodynamic,saryal2021PRE,rignon2021PRE}.   
The TUR bound for an observable, such as the integrated charge current in quantum transport, was found smaller than the value of the original TUR over a certain parameter range \cite{ptaszynski_coherence-enhanced_2018,agarwalla_assessing_2018,liu2019thermodynamic,saryal2021PRE}. 
In these studies, the quantum nature was suspected to enhance the precision of the current, and thus lowers the TUR bound; however, expressed in the language of the transmission function of the nonequilibrium Green's function formalism, the physical origin of the loose TUR bound
was left unclear  \cite{ptaszynski_coherence-enhanced_2018,agarwalla_assessing_2018,liu2019thermodynamic,saryal2021PRE}.  

\begin{figure}[t]
\includegraphics[width=0.8\columnwidth]{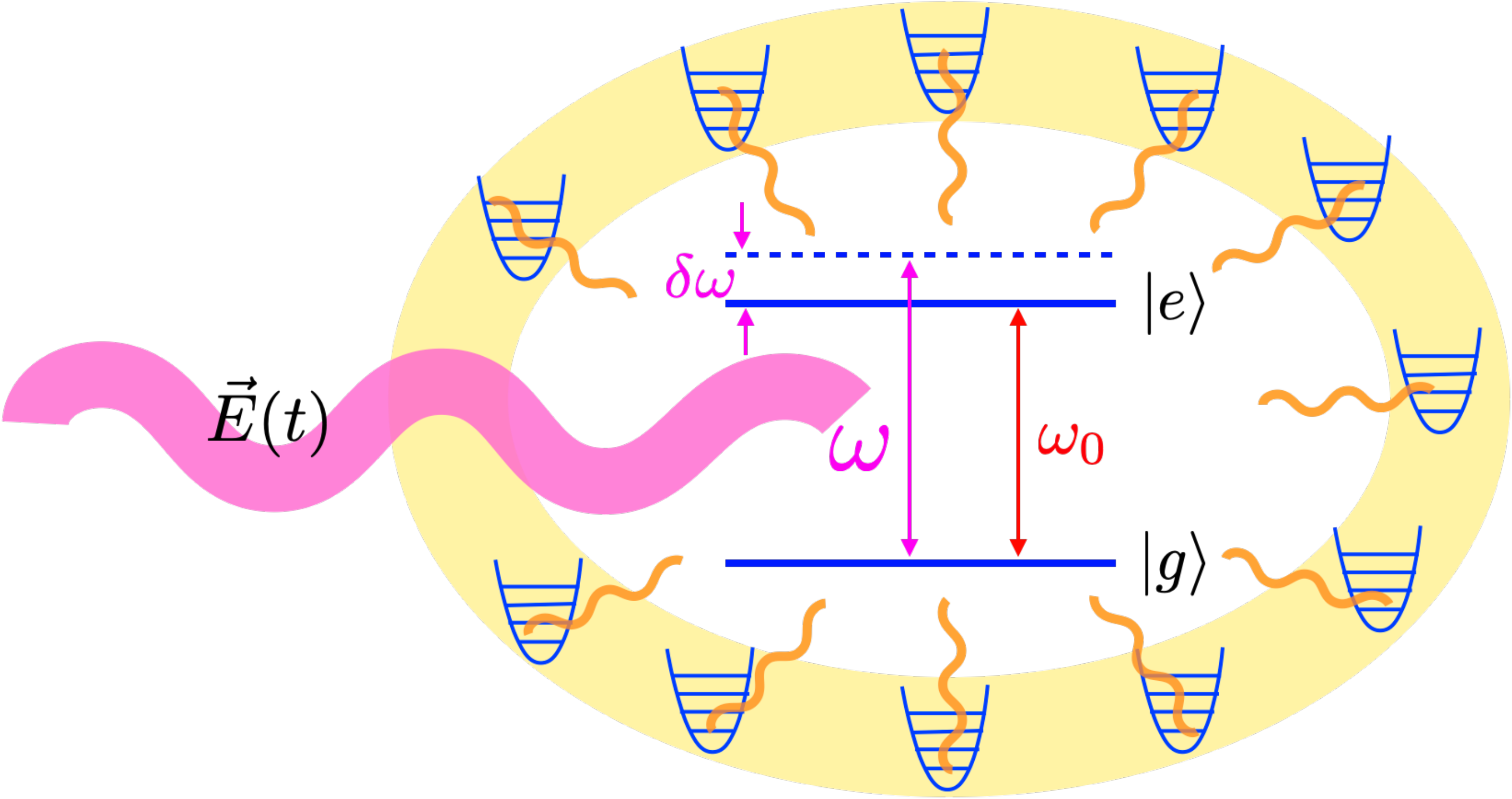}
\caption{A non-degenerate TLS coupled to bosonic bath, perturbed by the incident field with the angular frequency $\omega$. The resonant frequency is denoted by $\omega_0$. 
The difference between $\omega$ and $\omega_0$, $\delta\omega = \omega - \omega_0$, is the detuning. 
}
\label{driven_two_level}
\end{figure}

Here we consider a simple example of open quantum process which allows us to dissect the physical origin of the loose TUR bound. 
The model is a standard two-level system (TLS) of ground ($\ket{g}$) and excited ($\ket{e}$) states coupled to bosonic bath or radiation field maintained at temperature $T$. 
The transitions between the two levels are stimulated by an external driving field via dipole-electric field interaction \cite{Breuer,Carmichael,Leggett1987RMP} (Fig.~\ref{driven_two_level}). 
Without driving, it is expected that the absorption ($\ket{g}\rightarrow \ket{e}$) and emission ($\ket{e}\rightarrow\ket{g}$) satisfy the detailed balance (DB) condition with zero mean current ($\langle j\rangle=0$), and in this case the population ratio of the ground and excited states must maintain the canonical distribution, as a result of exchanging energy with the thermal bath via system-bath coupling. 
On the other hand, the external field, perturbing the system via the dipole-field coupling ($-\vec{d}\cdot\vec{E}$), generates a transition current ($\langle j\rangle>0$) between the two levels, breaking the DB condition, which generates heat current dissipating to the bath.
The excess number of transitions stimulated by the field 
is countered by the bath that invokes time irreversible relaxation of the system to its NESS.  

The expressions for the mean current of net transitions, current fluctuations, and entropy production of the model enable us to clarify the detail of quantum nature contributing to loosening the TUR bound.

\section{Model} 
The total Hamiltonian in the presence of an external radiation field is given by \cite{Carmichael,scully_zubairy_1997}
\begin{align}
H(t)&=H_S+H_\text{ext}(t)+H_B+H_{SB}
\end{align}
with  
\begin{align}
H_S&= (\hbar/2)\omega_0\sigma_{z} \nonumber\\
H_\text{ext}(t)&=-  \vec{d}\cdot\vec{E}({\bf r},t) \nonumber\\
H_B&= \sum_{\textbf{k},\xi} \hbar\omega_{\textbf{k}}b_{\textbf{k},\xi}^{\dagger}b_{\textbf{k},\xi} \nonumber\\
H_{SB}&=\sum_{\textbf{k},\xi}\hbar\left(g_{\textbf{k},\xi}^{\ast}b_{\textbf{k},\xi}^{\dagger}\sigma_{-} + g_{\textbf{k},\xi}b_{\textbf{k},\xi}\sigma_{+}\right)\nonumber.
\end{align}
Here $\sigma$'s are the Pauli spin operators: 
$\sigma_z = \ket{e}\bra{e} - \ket{g}\bra{g}$, 
$\sigma_+ = \ket{e}\bra{g}$, 
and $\sigma_- = \ket{g}\bra{e}$. 
$H_{\rm ext}(t)$ is defined by the dipole operator $\vec{d}= \vec{d}_{eg}\sigma_+ + \vec{d}_{ge}\sigma_-$ with 
the transition dipole $\vec{d}_{eg}$ ($\vec{d}_{ge}$) from $\ket{g}$ to $\ket{e}$ (from $\ket{e}$ to $\ket{g}$), 
interacting with the electric field 
$\vec{E}({\bf r},t)=\vec{\varepsilon}e^{-i({\bf k}\cdot{\bf r}-i\omega t)}+\vec{\varepsilon}^\ast e^{i({\bf k}\cdot{\bf r}-i\omega t)}$, where ${\bf k}$ is the wavevector interacting with the system at ${\bf r}$, and $\omega$ is the angular frequency of the field. In $H_B$ and $H_{SB}$,  
the summation $\sum_{\textbf{k},\xi}$ runs over the wavevector $\textbf{k}$ and polarization $\xi$. 
The symbols, $b_{\textbf{k},\xi}^{\dagger}$ and $b_{\textbf{k},\xi}$ are the creation and annihilation operators for the bath modes represented by the harmonic oscillators with angular frequency $\omega_k$. 
The strength of system-bath coupling is quantified by $g_{\textbf{k},\xi}\equiv -ie^{i{\bf k}\cdot{\bf r}}\sqrt{2\pi\omega_{\bf k}/\hbar V}\hat{\varepsilon}_{\textbf{k},\xi}\cdot\vec{d}_{eg}$, which is defined in terms of the polarization vector of electric field $\hat{\varepsilon}_{\textbf{k},\xi}$, quantization volume $V$.

The total density matrix $\rho_\text{tot}(t)$ undergoes a unitary evolution,  
obeying the von Neumann equation, 
\begin{align}\label{Von_Neumann_Sch}
\frac{d\rho_\text{tot}(t)}{dt}=-\frac{i}{\hbar}\left[H(t),\rho_\text{tot}(t)\right].
\end{align}  
Next, tracing out the bath degrees of freedom from the total density matrix, we obtain the evolution equation of reduced density matrix for the system 
$\rho(t)= \text{Tr}_B(\rho_\text{tot}(t))$, by assuming that $\rho_\text{tot}(t)$ can be described with the uncorrelated product state of the system $\rho(t)$ and environment in equilibrium ($\rho_B^{eq}$), namely $\rho_\text{tot}(t)\approx \rho(t)\otimes\rho_B^{eq}$. 

We employ the Lindblad approach \cite{Carmichael,Breuer} (see Appendix A for details) which ensures that the density matrix is self-adjoint ($\rho_{eg}=\rho_{ge}^*$), trace-preserving ($\rho_{ee}+\rho_{gg}=1$), and positive semi-definite, while the system evolves and relaxes to the steady state.  
The evolution equation for $\rho(t)$ reads
\begin{align}
\frac{d\rho(t)}{dt} &= - \frac{i}{\hbar}[H_{S}+H_\text{ext}(t), \rho(t)] +\mathcal{D}(\rho(t)).  
\label{master}
\end{align}
In evaluating the effect of $H_\text{ext}(t)$ on $\rho(t)$, 
we consider the dipole approximation that the driving field is nearly constant over the molecular scale ($e^{i{\bf k}\cdot{\bf r}}\simeq 1$) \cite{scully_zubairy_1997}, such that $\vec{E}(t)\simeq \vec{\varepsilon}e^{i\omega t}+\vec{\varepsilon}^\ast e^{-i\omega t}$.  Together with $\vec{d}= \vec{d}_{eg}\sigma_+ + \vec{d}_{ge}\sigma_-$, one obtains   
$H_\text{ext}(t)=-\hbar\Omega(e^{i\omega t}+e^{-i\omega t})(\sigma_++\sigma_-)$ with the \emph{driving frequency}, $\Omega=\vec{d}_{eg}\cdot\vec{\varepsilon}/\hbar=\vec{d}_{ge} \cdot\vec{\varepsilon}^\ast/\hbar$. 
The range of the angular frequency $\omega$ relevant to our model is $\omega=\omega_0+\delta\omega$ with $|\delta\omega|/\omega_0\ll 1$, where $\delta\omega$ is the detuning that reflects the extent of off-resonance (see Fig.~\ref{driven_two_level}). 
The time-irreversible dynamics of system, invoked by a weak system-bath coupling ($\omega\gg \gamma$), is delineated by the Lindblad \emph{dissipator} (see Appendix A for the derivation),   
\begin{align}
\mathcal{D}(\rho(t))
&\equiv 
\gamma(\bar{n}+1)\left(\sigma_-\rho(t)\sigma_+ - \frac{1}{2}\Big\{\sigma_+\sigma_-,\rho(t)\Big\}\right)\nonumber\\
&+\gamma\bar{n}\left(\sigma_+\rho(t)\sigma_- - \frac{1}{2}\Big\{\sigma_-\sigma_+,\rho(t)\Big\}\right),  
\label{eqn:Lindblad}
\end{align}
where $\bar{n}=(e^{\beta\hbar\omega_0} - 1)^{-1}$ denotes the mean occupation number of bosonic bath, and $\{\hat{A},\hat{B}\}=\hat{A}\hat{B}+\hat{B}\hat{A}$ is the anti-commutator.  
The parameter $\gamma$, involving the spontaneous emission frequency, sets the timescale of relaxation of the system to the steady state, and $\gamma\bar{n}$ represents the rate of bath-induced transition.  
Thus, the first and second terms of Eq.~\ref{eqn:Lindblad} arise from emissions and absorptions, respectively. 

Eqs.~\ref{master} and \ref{eqn:Lindblad} yield 
the evolution equations for the reduced density matrix elements as follows
\begin{align}
\dot{\rho}_{ee} =&-\gamma(\bar{n} + 1)\rho_{ee} -i\Omega\left( e^{i\omega t} + e^{-i\omega t} \right)\rho_{eg}\nonumber\\
&+ i\Omega\left( e^{i\omega t} + e^{-i\omega t} \right)\rho_{ge} + \gamma\bar{n}\rho_{gg}, \nonumber\\
\dot{\rho}_{eg} =&-i\Omega\left( e^{i\omega t} + e^{-i\omega t} \right)\rho_{ee} - \left[i\omega_0 +\frac{\gamma}{2}(2\bar{n} + 1) \right]\rho_{eg} \nonumber\\
&+ i\Omega\left( e^{i\omega t} + e^{-i\omega t} \right)\rho_{gg}\nonumber\\
\dot{\rho}_{ge} =&i\Omega\left( e^{i\omega t} + e^{-i\omega t} \right)\rho_{ee} + \left[i\omega_0 -\frac{\gamma}{2}(2\bar{n} + 1) \right]\rho_{ge} \nonumber\\
&- i\Omega\left( e^{i\omega t} + e^{-i\omega t} \right)\rho_{gg}\nonumber\\
\dot{\rho}_{gg} =&\gamma(\bar{n} + 1)\rho_{ee} +i\Omega\left( e^{i\omega t} + e^{-i\omega t} \right)\rho_{eg}\nonumber\\
&- i\Omega\left( e^{i\omega t} + e^{-i\omega t} \right)\rho_{ge} - \gamma\bar{n}\rho_{gg}
\label{dynamical_equation1}
\end{align}
Transformation of the density matrix into the one in the rotating frame, 
$\rho_{ee} \rightarrow \tilde{\rho}_{ee}$, $\rho_{gg} \rightarrow \tilde{\rho}_{gg}$, and 
$\rho_{eg} \rightarrow \tilde{\rho}_{eg}e^{-i\omega t}$ (see Appendix B) produces the terms retaining $(1+e^{\pm2i\omega t})$ in Eq.~\ref{dynamical_equation1}. 
With the \emph{rotating wave approximation} (RWA), the highly oscillatory terms can be ignored in comparison with the driving frequency $\Omega$, which approximates $(1+e^{\pm2i\omega t})$ to 1. To be specific, the term $e^{\pm2i\omega t}$ vanishes when it is averaged over a time interval $\sim\Omega^{-1}$ with the condition of $\omega/\Omega\gg 1$: 
\begin{align}
\Big|\frac{1}{\Omega^{-1}}\int_{-\frac{1}{2\Omega}}^{\frac{1}{2\Omega}}e^{\pm2i\omega t}dt\Big|=\Big|\frac{\sin{(\omega/\Omega)}}{\omega/\Omega}\Big|\leq \Omega/\omega\ll 1. 
\label{eqn:RWA}
\end{align} 
Thus, the RWA simplifies Eq.~\ref{dynamical_equation1} to 
\begin{align}
\frac{\partial \tilde{\rho}_{ee}}{\partial \tau} =&-(\bar{n} + 1)\tilde{\rho}_{ee} -i\Omega_\gamma\tilde{\rho}_{eg} + i\Omega_\gamma\tilde{\rho}_{ge} + \bar{n}\tilde{\rho}_{gg}  \nonumber\\
\frac{\partial \tilde{\rho}_{eg}}{\partial \tau} =&-i\Omega_\gamma\tilde{\rho}_{ee} + \left[i\delta\omega_\gamma -\frac{(2\bar{n} + 1)}{2} \right]\tilde{\rho}_{eg} + i\Omega_\gamma\tilde{\rho}_{gg}\nonumber\\
\frac{\partial \tilde{\rho}_{ge}}{\partial \tau} =&i\Omega_\gamma\tilde{\rho}_{ee} - \left[i\delta\omega_\gamma +\frac{(2\bar{n} + 1)}{2} \right]\tilde{\rho}_{ge} - i\Omega_\gamma\tilde{\rho}_{gg}\nonumber\\
\frac{\partial \tilde{\rho}_{gg}}{\partial \tau} =&(\bar{n} + 1)\tilde{\rho}_{ee} +i\Omega_\gamma\tilde{\rho}_{eg} - i\Omega_\gamma\tilde{\rho}_{ge} - \bar{n}\tilde{\rho}_{gg}
\label{dynamical_equation2}
\end{align}
where the whole equation is rescaled with $\gamma^{-1}$, with the definitions of dimensionless parameters $\tau\equiv \gamma t$, 
$\Omega_\gamma\equiv \Omega/\gamma$, and $\delta\omega_\gamma \equiv \delta\omega/\gamma$. 
It is noteworthy that once the RWA is taken under the condition of $\omega/\Omega\gg1$, the dynamics of our TLS model studied in the rotating frame with the angular frequency $\omega$ depends only on $\delta\omega_\gamma$, and is impervious to the frequency ($\omega$) of the driving field ($\vec{E}(t)$) (see Eq.~\ref{dynamical_equation2}).

The steady state values of density matrix elements are obtained from Eq.~\ref{dynamical_equation2} by setting $d\tilde{\rho}/d\tau=0$ with the constraints of $\tilde{\rho}_{ee}^{ss}+\tilde{\rho}_{gg}^{ss}=1$ and $\tilde{\rho}_{eg}^{ss}=(\tilde{\rho}_{ge}^{ss})^\ast$ 
\begin{align}
&\tilde{\rho}_{ee}^{ss}=\dfrac{\bar{n}[(2\bar{n} + 1)^2+4\delta\omega_\gamma^2] + 4\Omega_\gamma^2(2\bar{n}+1)}{(2\bar{n} + 1)[(2\bar{n} + 1)^2 + 4\delta\omega_\gamma^2 + 8\Omega_\gamma^2]} \nonumber\\
&\tilde{\rho}_{gg}^{ss}=\dfrac{(\bar{n}+1)[(2\bar{n} + 1)^2+4\delta\omega_\gamma^2] + 4\Omega_\gamma^2(2\bar{n}+1)}{(2\bar{n} + 1)[(2\bar{n} + 1)^2 + 4\delta\omega_\gamma^2 + 8\Omega_\gamma^2]}\nonumber\\
&\tilde{\rho}_{eg}^{ss}=\frac{-2\Omega_\gamma\left[2\delta\omega_\gamma - i(2\bar{n} + 1)\right]}{(2\bar{n} + 1)[(2\bar{n} + 1)^2 + 4\delta\omega_\gamma^2 + 8\Omega_\gamma^2]}. 
\label{eqn:ss}
\end{align}
For $\Omega_\gamma=0$, the TLS coupled only to the thermal bath should be at equilibrium, satisfying the DB condition \cite{tscherbul_long-lived_2014}, which can be confirmed from 
Eq.~\ref{eqn:ss}. 
The ratio of excited and ground state populations obeys the Boltzmann distribution, $\tilde{\rho}^{ss}_{ee}/\tilde{\rho}^{ss}_{gg} = \bar{n}/(\bar{n}+1)=e^{-\beta\hbar\omega_0}$, 
and the quantum coherence vanishes ($\tilde{\rho}_{eg}^{ss} = 0$). 
In contrast, the model with $\Omega_\gamma\neq 0$ results in the breakdown of the DB condition and nonvanishing coherence ($\tilde{\rho}_{eg}^{ss} \neq 0$). 

We introduce simplified notations for the real ($\rho_R\equiv \text{Re}(\tilde{\rho}_{eg}^{ss})$) and imaginary parts ($\rho_I\equiv \text{Im}(\tilde{\rho}_{eg}^{ss})$) of the coherence ($\tilde{\rho}^{ss}_{eg}=\rho_R+i\rho_I$):  
\begin{align}
\rho_R&\equiv\frac{-4\Omega_\gamma\delta\omega_\gamma}{\coth{\frac{\mathcal{A}}{2}}[\coth^2{\frac{\mathcal{A}}{2}}+4\delta\omega_\gamma^2+8\Omega_\gamma^2]}\nonumber\\
\rho_I&\equiv\frac{2\Omega_\gamma}{[\coth^2{\frac{\mathcal{A}}{2}}+4\delta\omega_\gamma^2+8\Omega_\gamma^2]},  
\label{eq:coherence}
\end{align} 
where we have used $\coth{\frac{\mathcal{A}}{2}}=(2\bar{n}+1)$ with $k_BT\mathcal{A}\equiv \hbar\omega_0$ which corresponds to the energy gap between the ground and excited states.  The gain of energy eventually dissipates into the bath along a single cycle of absorption and emission.  
A couple of remarks on $\rho_R$ and $\rho_I$ are in place. 
(i) Both $\rho_R$ and $\rho_I$ vanish with the driving frequency as $\sim 1/\Omega_\gamma$. 
(ii) $\rho_I$ takes the form of Lorentzian with respect to $\delta\omega_\gamma$, whereas $\rho_R$ is an odd function of $\delta\omega_\gamma$. 
In fact, the quantum coherence can be related to the response function (or susceptibility) of the system between the external electric field ($\vec{E}$) and the polarization ($\vec{P}$) in the linear response regime ($\vec{P}=\chi\vec{E}$). More specifically, 
$\rho_R$ and $\rho_I$ are related to the real and imaginary parts of the susceptibility, which inform about the dispersion and absorption profile of the light-matter interaction, respectively (see Appendix C).

\section{Mean current and current fluctuations}
The mean current and the current fluctuations involving the net number of transitions, $n(\tau)$, i.e., the difference between the total numbers of emissions and absorptions for time $\tau$, at steady states, elicited by the irradiation are calculated 
by employing the method of generating function \cite{Koza1999JPA,bruderer_inverse_2014} (see Appendix D): 
\begin{align}
\langle j\rangle &\equiv \lim_{\tau\rightarrow\infty}\frac{\langle n(\tau)\rangle}{\tau}=\frac{4\Omega_\gamma^2}{\left[\coth^2{\frac{\mathcal{A}}{2}} + 4\delta\omega_\gamma^2 + 8\Omega_\gamma^2 \right] }\nonumber\\
 &=2\Omega_\gamma\rho_I, 
\label{eqn:j}
\end{align}
and 
\begin{align}
\text{Var}[j]&\equiv\lim_{\tau\rightarrow\infty}\frac{\text{Var}[n(\tau)]}{\tau}\nonumber\\
&=\langle j\rangle\coth{\left(\frac{\mathcal{A}}{2}\right)}f(\mathcal{A},\Omega_\gamma,\delta\omega_\gamma)
\label{cumm}
\end{align}
with 
\begin{align}
&f(\mathcal{A},\Omega_\gamma,\delta\omega_\gamma)\nonumber\\
&\equiv \left(1 + \dfrac{[32\delta\omega_\gamma^2-24\coth^2{\frac{\mathcal{A}}{2}}]\Omega_\gamma^2}{\coth^2{\frac{\mathcal{A}}{2}}\left[\coth^2{\frac{\mathcal{A}}{2}} + 4\delta\omega_\gamma^2 + 8\Omega_\gamma^2 \right]^2}\right)\nonumber\\
&=\left[1+2\rho_R^2-6\rho_I^2\right]. 
\label{eqn:f}
\end{align}
The expression of mean current (Eq.~\ref{eqn:j}) conforms to the \emph{coherence-current relation} \cite{yang_steady-state_2020,wu_efficient_2012} which states that   
the imaginary part of the coherence (Eq.~\ref{eq:coherence}) is responsible for inducing a quantum current between the states. 
Notice that $\langle j\rangle\geq 0$ for the entire parameter space with $\langle j\rangle=0$ realized when the driving field is absent ($\Omega_\gamma=0$) or the energy gap (dissipation to the bath) is zero ($\mathcal{A}=0$) and that $\langle j\rangle$ increases monotonically with $\Omega_\gamma$ and upper-bounded to $\langle j\rangle\rightarrow 1/2$ for $\Omega_\gamma\gg 1$ ($y\neq 0$). 
Next, Eq.~\ref{cumm} clarifies that $\text{Var}[j]=0$ when $\Omega_\gamma=0$. 
For $\Omega_\gamma\neq 0$ and $|\delta\omega_\gamma|\gg 1$, the current fluctuations are upper bounded as $\text{Var}[j]\leq (1/2)\coth{(\frac{\mathcal{A}}{2})}$; and for $\delta\omega_\gamma=0$, it is simplified to $\text{Var}[j]/\langle j\rangle=\coth{(\frac{\mathcal{A}}{2})}(1-6\rho_I^2)$, where $\rho_I=\rho_I(\Omega_\gamma)$ (Eq.\ref{eq:coherence}) is a non-monotonic function of $\Omega_\gamma$. 

\section{Steady-state entropy production rate}
The presence of irreversible current is conducive to the entropy production. 
The total entropy production rate in the open quantum system is given by \cite{Breuer,breuer_quantum_2003}
\begin{align}
\Sigma &= \dfrac{dS}{d\tau} + J_S. 
\label{entropy1}
\end{align}
The entropy of the system is defined with the von Neumann entropy, $S(\tilde{\rho}) = -k_B\text{Tr}(\tilde{\rho}\log{\tilde{\rho}})$, and $J_S$ corresponds to the heat current into the bath. 
At steady state ($\dot{\tilde{\rho}} = 0$), $dS(\tilde{\rho})/d\tau = 0$, so that 
$\Sigma$ is contributed by $J_S$ part only.   
In the Lindblad framework, $J_S$ is expressed as \cite{Breuer,breuer_quantum_2003}
$J_S = (\hbar\omega_0/T)\text{Tr} \left[ (\bar{n} + 1)\sigma_+\sigma_-\tilde{\rho} - \bar{n}\sigma_-\sigma_+\tilde{\rho} \right]$.
Therefore, the total entropy production rate at steady states is 
\begin{align}
\Sigma^{ss}/k_B &= \beta\hbar\omega_0\text{Tr} \left[ (\bar{n} + 1)\sigma_+\sigma_-\tilde{\rho}^{ss} - \bar{n}\sigma_-\sigma_+\tilde{\rho}^{ss} \right] \nonumber\\
& = \beta\hbar\omega_0 \left[ (\bar{n} + 1)\tilde{\rho}^{ss}_{ee} - \bar{n}\tilde{\rho}^{ss}_{gg} \right]  \nonumber\\
& = \beta\hbar\omega_0\dfrac{4\Omega_\gamma^2}{\left[ (2\bar{n} + 1)^2 + 4\delta\omega_\gamma^2 + 8\Omega_\gamma^2 \right]}\nonumber\\
&=\mathcal{A}\times\langle j\rangle\geq 0. 
\label{entropy3}
\end{align}
Despite $dS(\tilde{\rho})/d\tau=0$ at NESS, the non-vanishing heat current due to the interaction with the external field 
($J_S^{ss}\neq0$) gives rise to $\Sigma^{ss}\neq 0$. 
We also note that the value of $\Sigma^{ss}$ is unaffected by the transformation into the rotating frame $\tilde{\rho}_{ij}=e^{i\omega\sigma_zt/2}\rho_{ij}e^{-i\omega\sigma_zt/2}$. 

\begin{figure}[t]
\includegraphics[width=1.0\linewidth]{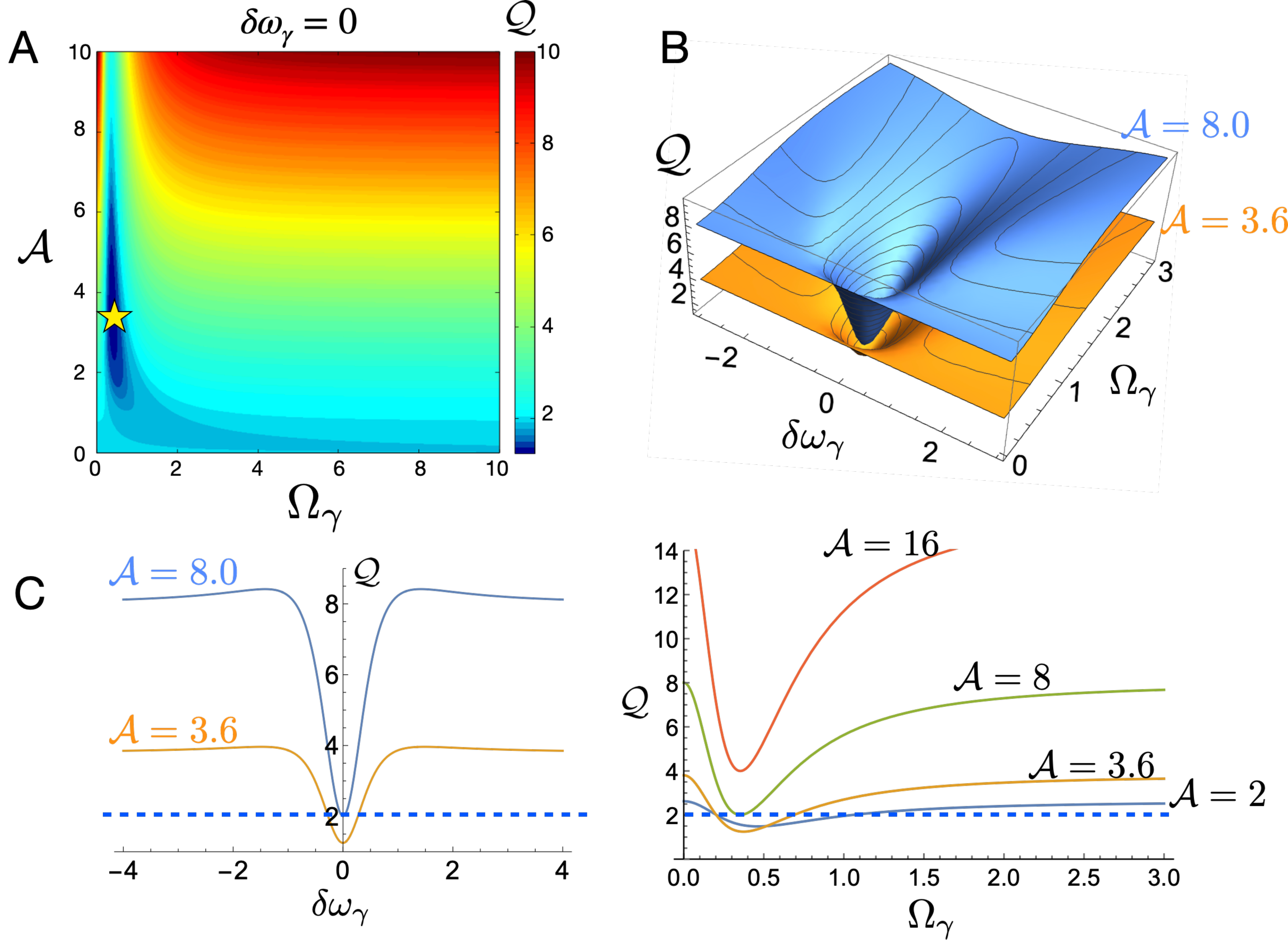}
\caption{The uncertainty product $\mathcal{Q}$. 
(A) $\mathcal{Q}$ as a function of $\Omega_\gamma$ and $\mathcal{A}$ for the case of perfect resonance ($\delta\omega_\gamma = 0$).  
The condition of $(\Omega_\gamma,\mathcal{A})$ that minimizes $\mathcal{Q}$ is marked with the star symbol. 
(B) The diagram of $\mathcal{Q}=\mathcal{Q}(\Omega_\gamma,\delta\omega_\gamma)$ for $\mathcal{A}=3.6$ and $\mathcal{A}=8$. 
(C) Dissections of $\mathcal{Q}$. (left) $\mathcal{Q}(\Omega_\gamma=0.37,\delta\omega_\gamma)$ at $\mathcal{A}=3.8$ and 8.  (right) $\mathcal{Q}(\Omega_\gamma,\delta\omega_\gamma=0)$ for various values of $\mathcal{A}$. 
The bound of the original TUR ($\mathcal{Q}=2$) is marked with the blue dashed line. 
}
\label{Resonant}
\end{figure}

\section{Thermodynamic uncertainty relation}
The uncertainty product of TUR at NESS \cite{barato2015PRL,horowitz_thermodynamic_2020} for the net transitions of the TLS is written as 
\begin{align}
\mathcal{Q}&=\lim_{\tau\rightarrow\infty}\dfrac{\Sigma^{ss}\tau}{k_B}\dfrac{\langle \delta n(\tau)^2\rangle}{\langle n(\tau)\rangle^2}=\underbrace{\left(\dfrac{\Sigma^{ss}}{k_B\langle j\rangle}\right)}_{=\mathcal{A}}\underbrace{\left(\dfrac{\text{Var}[j]}{\langle j\rangle}\right)}_{=\mathcal{F}}\nonumber\\
&=\mathcal{A}\coth{\left(\frac{\mathcal{A}}{2}\right)}
f(\mathcal{A},\Omega_\gamma,\delta\omega_\gamma). 
\label{TUR}
\end{align}
The original TUR derived for classical system 
states that the uncertainty product $\mathcal{Q}$ between the cost ($\Sigma^{ss}/k_B$) and the precision dictated by the square of the relative fluctuations of current ($\text{Var}[j]/\langle j\rangle^2$) cannot be smaller than $2$ \cite{barato2015PRL,gingrich_dissipation_2016,horowitz_thermodynamic_2020}. 
However, when TUR is extended to quantum systems, the uncertainty product can have a loose bound. 
The expression given in Eq.~\ref{eqn:f} points to the possibility of $f(\mathcal{A},\Omega_\gamma,\delta\omega_\gamma)<1$, which in turn leads to $\mathcal{Q}<2$, when 
the imaginary part of coherence is dominant.  
In fact, $\rho_R=0$ for the case of perfect resonance ($\delta\omega_\gamma=0$)  
simplifies $\mathcal{Q}$ to $\mathcal{Q}(\mathcal{A},\Omega_\gamma)=
\mathcal{A}\coth{(\frac{\mathcal{A}}{2})}\left[1-\frac{24\Omega_\gamma^2}{(\coth^2{(\frac{\mathcal{A}}{2})}+8\Omega_\gamma^2)^2}\right]\geq\mathcal{Q}(\mathcal{A}^\ast,\Omega_\gamma^\ast)\approx 1.25$ at $(\mathcal{A}^\ast,\Omega_\gamma^\ast)\approx(3.61,0.37)$ (see Fig.~\ref{Resonant}A). 
The minimal uncertainty product, $\mathcal{Q}\geq\mathcal{Q}_{\rm min}=1.25$ coincides with the value obtained in a recent work that has studied TUR of a spin in a rotating magnetic field \cite{menczel2021thermodynamic}. 

For the case of finite detuning (off-resonance) $\delta\omega_\gamma\neq 0$, 
the real part of coherence ($\rho_R\neq0$) contributes to increasing the size of $\mathcal{Q}$ (Fig.~\ref{Resonant}B, C). 
Thus, for a given set of parameter $(\mathcal{A},\Omega_\gamma,\delta\omega_\gamma)$, the value of $\mathcal{Q}$ is determined 
as a result of competition between the real and imaginary parts of the coherence. 
Notably, $\rho_R^2<3\rho_I^2$, i.e., $|\delta\omega_\gamma|<(\sqrt{3}/2)\coth{\frac{\mathcal{A}}{2}}$, is a necessary, but not a sufficient condition for the loose bound of TUR. 

TUR can also be recast into the form of an inequality of the Fano factor of transitions \cite{hasegawa2019PRL}. 
For our system,  
Eqs.~\ref{eqn:j} and ~\ref{cumm} define the Fano factor of transitions, 
\begin{align}
\mathcal{F}&=\lim_{\tau\rightarrow\infty}\frac{\text{Var}[n(\tau)]}{\langle n(\tau)\rangle}=\frac{\text{Var}[j]}{\langle j\rangle}\nonumber\\
&=\coth{(\frac{\mathcal{A}}{2})}f(\mathcal{A},\Omega_\gamma,\delta\omega_\gamma)  
\geq \phi(\mathcal{A}). 
\label{eqn:Fano}
\end{align}
For the original TUR derived for the case of constant driving \cite{barato2015PRL,gingrich_dissipation_2016,horowitz_thermodynamic_2020}, $\phi(\mathcal{A})=\phi_o(\mathcal{A})=2/\mathcal{A}$. 
For the case of \emph{time-symmetric periodic driving}, Proesmans and Van den Broeck have shown that the steady state current fluctuations per period in the long-time limit is bounded by $
\text{Var}[j]/\langle j\rangle^2\geq 2\Delta t/(e^{\Sigma^{ss}\Delta t/k_B}-1)$ \cite{Proesmans:2017,horowitz_thermodynamic_2020}. 
Our electromagnetic field-driven quantum TLS is, in principle, under time-symmetric periodic driving with a constant period of $\Delta t=2\pi/\omega$;  however, such a periodicity in driving field under weak field limit ($\Omega\ll \omega$) is reduced to a constant driving under RWA, as detailed in Eqs.~\ref{eqn:RWA} and \ref{dynamical_equation2}. 
Thus, it is pertinent to ask whether or not the uncertainty product of our quantum TLS satisfies the inequality dictated by the classical version of TUR under a constant driving.  
Since $\langle j\rangle \Delta t=1$ from the definition given in Eq.~\ref{eqn:j} and $\Sigma^{ss}\Delta t=\mathcal{A}$, the expression can be cast into $\text{Var}[j]/\langle j\rangle \geq \phi(\mathcal{A})=\phi_p(\mathcal{A})=2/(e^{\mathcal{A}}-1)$. 
Mathematically, $\phi_o(\mathcal{A})>\phi_p(\mathcal{A})$ for all $\mathcal{A}\geq 0$, but $\phi_p(\mathcal{A})$ converges to the form of $\phi_o(\mathcal{A})$ when $\mathcal{A}\ll 1$. 

\begin{figure}[t]
\includegraphics[width=0.6\columnwidth]{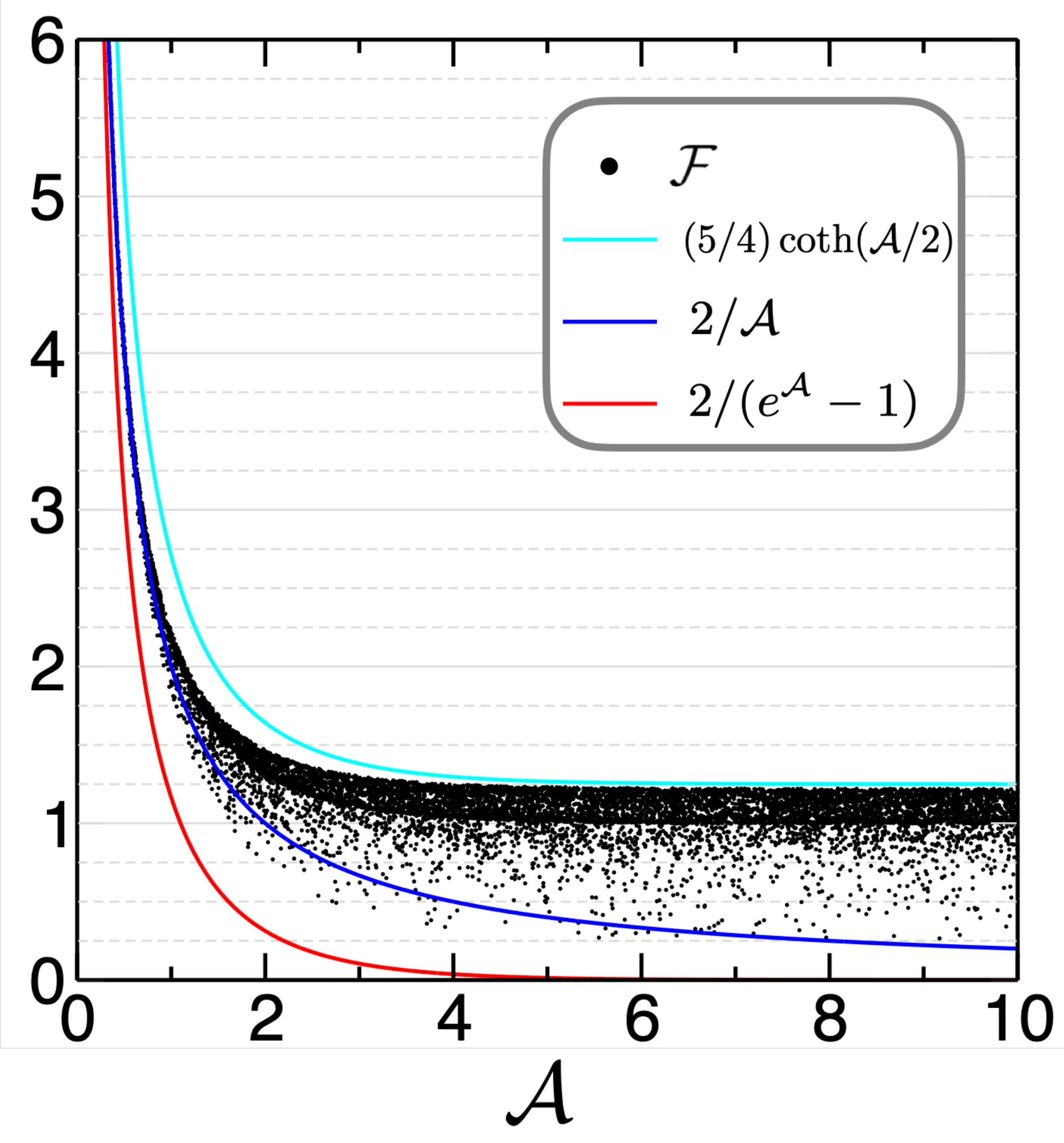}
\caption{Fano factor ($\mathcal{F}$) of a bath-coupled TLS under a weak driving plotted against the net entropy production over a cycle of absorption and emission, $\mathcal{A}(=\beta\hbar\omega_0)$. 
For $\mathcal{F}$ calculated with a randomly generated set of parameters $(\mathcal{A},\Omega_\gamma,\delta\omega_\gamma)$, 
there are cases whose Fano factor is smaller than the lower bound of the original TUR (blue), but still greater than the bound specified by the generalized uncertainty relation (red) \cite{Proesmans:2017}. 
}
\label{lower_bound}
\end{figure}

$\mathcal{F}$, plotted against $\mathcal{A}$ with randomly varying set of parameters $(\mathcal{A},\Omega_\gamma,\delta\omega_\gamma)$, is upper bounded by $(5/4)\coth{\frac{\mathcal{A}}{2}}$ (the cyan line in Fig.~\ref{lower_bound}), which is obtained from 
$f(\mathcal{A},\Omega_\gamma,\delta\omega_\gamma)\approx 1+\frac{32\delta\omega_\gamma^2\Omega_\gamma^2}{[4\delta\omega_\gamma^2+8\Omega_\gamma^2]^2}\leq 5/4$ for $\mathcal{A}\gg 1$, $|\delta\omega_\gamma|\gg 1$ and $\Omega_\gamma\gg 1$. 
Of particular note is that there are data points that satisfy $\phi_p(\mathcal{A})<\mathcal{F}(\mathcal{A},\Omega_\gamma,\delta\omega_\gamma)<\phi_o(\mathcal{A})$ at $1\lesssim\mathcal{A}\lesssim6$, signifying that the TUR of our quantum TLS is characterized with the bound less tight than that of the original TUR (Fig.~\ref{lower_bound}).

\section{Discussions}
The relation between the parameters $\omega$, $\omega_0$, $\gamma$, and $\Omega$,  
\begin{align}
\omega\sim \omega_0\gg \gamma\sim \Omega    
\label{eqn:relation}
\end{align}
is the essential condition to study the time-irreversible evolution of our TLS. 
First, for transitions induced by dipole-electric field coupling, 
Fermi's golden rule offers an expression for the emission rate, $\gamma =4\omega_0^3| d_{eg}|^2/(3\hbar c^3)$ \cite{Breuer}. 
Provided that the dipole moment of a molecule (or an atom) is given by $|d_{eg}|=ea_0$ with its 
electric charge $e$ and size $a_0(\lesssim 1$ nm), and that $\omega_0/c=2\pi/\lambda$ where $c$ is the speed of light, and $\lambda(\approx 500$ nm) is the wavelength of the field in the optical domain, one gets $\omega_0/\gamma\approx 6.5\times 10^5$ from 
$\gamma=(4\alpha/3)\left(2\pi a_0/\lambda\right)^2\omega_0$, with 
$\alpha\equiv e^2/\hbar c=1/137$ and $2\pi a_0/\lambda\sim 10^{-2}$. 
The separation of timescales ($\omega/\gamma\gg 1$) justifies decomposition of the total density matrix into the system and equilibrated bath parts, 
and rationalizes the truncation of quantum optical master equation at the second order (\emph{Born-Markov approximation}. See Appendix A). 
Second, the condition of $\omega\gg\Omega$ was used in taking the RWA in Eq.~\ref{eqn:RWA} to obtain Eq.~ \ref{dynamical_equation2}. 

Since the Eq.~\ref{dynamical_equation2} is independent of $\omega$, 
the system is effectively under a constant driving. 
As a result, the system at long time limit, if classical, should obey the original TUR \cite{barato2015PRL,fischer2020PRE}, instead of the generalized TUR under time-symmetric periodic driving \cite{Proesmans:2017}. 
Thus, the loose TUR bound of our TLS model ($\mathcal{Q}<2$ in Fig.\ref{Resonant}) stems from the quantum nature of the dynamics.  
Our model further elaborates the contribution from the quantum nature to TUR:  
The imaginary part of quantum coherence contributes to suppressing the Fano factor of net transitions and lowering the TUR bound, whereas the real part of the coherence always tightens the TUR bound. 
The bound of TUR is determined by the competition between the dissipation and dispersion of the input light source. 

For the cases the system-bath coupling ($\gamma$) and/or the strength of driving field ($\Omega$) are strong and become comparable to the energy scale of the system, 
one should reevaluate the quantum master equation presented here based on generalized versions of open quantum dynamics that are under active development in recent years \cite{thingna2012JCP,tanimura2020JCP,ikeda2020ScienceAdvances}. 
However, as long as the energy gap of the TLS is in the optical domain ($\lambda\sim 500$ nm), one of the conditions identified for the loose TUR bound in Fig.~\ref{Resonant} ($\Omega_\gamma\lesssim 0.5$) ensures the relation specified in Eq.~\ref{eqn:relation}, which should reduce a generalized quantum master equation into the one studied here. 
Thus, we anticipate that the condition leading to the loose TUR bound will remain unaltered.  
Finally,  the other condition $\mathcal{A}\sim (2-8)$ (Fig.~\ref{Resonant})
becomes accessible in the range of high temperature $T=(2-8)^{-1}(hc/k_B\lambda)\sim \mathcal{O}(10^3)$ K for $\lambda\sim 500$ nm.  

Despite the apparent simplicity of the model presented here, it retains all the essential components that enables us to explore the TUR of quantum processes. 
The model also offers us a physically straightforward understanding of the condition that leads to loosening the TUR bound.    

\begin{acknowledgements}
This work was supported by the KIAS Individual Grants CG077602 (D.S.) and CG035003 (C.H.) from the Korea Institute for Advanced Study. 
We thank the Center for Advanced Computation in KIAS for providing computing resources.
\end{acknowledgements}

\section*{Appendix}

\setcounter{equation}{0}
\renewcommand{\theequation}{A\arabic{equation}}

\subsection{Derivation of Lindblad dissipator}

Introducing the transformations 
\begin{align}
H^{\prime}_{I}(t) &= e^{\frac{i}{\hbar}H_ot}H_{I}(t)e^{-\frac{i}{\hbar}H_ot}\nonumber\\
\rho^\prime_\text{tot}(t) &= e^{\frac{i}{\hbar}H_ot}\rho_\text{tot}(t)e^{-\frac{i}{\hbar}H_ot}
\end{align} 
with the definitions of 
\begin{align}
H_o&\equiv H_{S}+H_{B}\nonumber\\
H_I(t)&\equiv H_\text{ext}(t)+H_{SB},  
\end{align}
one can cast the von Neumann equation into the one in the interaction picture \cite{Carmichael}
\begin{align}
\frac{d\rho^\prime_\text{tot}(t)}{dt} = -\frac{i}{\hbar}[H^\prime_{I}(t),\rho^\prime_\text{tot}(t)]. 
\label{von_interaction}
\end{align}
After integrating Eq.(\ref{von_interaction}) and expanding it up to the second order, we obtain  
\begin{align}
\frac{d{\rho^\prime_\text{tot}}(t)}{dt} &=-\frac{i}{\hbar}[H^\prime_{SB}(t),\rho^\prime_\text{tot}(0)]-\frac{i}{\hbar}[H^\prime_{\text{ext}}(t),\rho^\prime_\text{tot}(0)] \nonumber\\
&-\frac{1}{\hbar^2}\int_0^tdt^{\prime}[H^\prime_{I}(t),[H^\prime_{I}(t^{\prime}),\rho^\prime_\text{tot}(t^{\prime})]]. 
\end{align}
Next, we trace out the bath modes, assuming that the density matrix $\rho_\text{tot}(t)$ for the total system at time $t$ can be described with the uncorrelated product state of system ($\rho(t)$) and environment in equilibrium ($\rho_{B}^{eq}$), such that 
$\rho_\text{tot}(t) \approx \rho(t) \otimes \rho_{B}^{eq}$. 
Then, it follows that  
\begin{widetext}
\begin{align}
\frac{d{\rho^\prime}(t)}{dt}= -\frac{i}{\hbar}[H^\prime_{\text{ext}}(t),\rho^\prime(0)]&-\frac{1}{\hbar^2}\int_{0}^{t} dt^{\prime} \Tr_B\left([H^\prime_{\text{ext}}(t),[H^\prime_{\text{ext}}(t^{\prime}),\rho^\prime(t^{\prime})\otimes \rho_{B}^{eq}]]\right)\nonumber\\
&-\frac{1}{\hbar^2}\int_{0}^{t} dt^{\prime} \Tr_B\left([H^\prime_{SB}(t),[H^\prime_{SB}(t^{\prime}),\rho^\prime(t^{\prime})\otimes \rho_{B}^{eq}]]\right), 
\label{master1}
\end{align}
\end{widetext}
where $\rho^\prime = \Tr_B\left(\rho^\prime_\text{tot}\right)$ was used in the left hand side of the equation and an assumption that the system is decoupled from the environment at $t=0$ leads to $\Tr_B[H^\prime_{SB}(t),\rho^\prime_\text{tot}(0)]=0$. 

Next, to proceed further, it is convenient to define $H_{SB} = \hbar\sum\limits_ms_mq_{m}$, where $s_m$ are the system operators and $q_{m}$ are the bath operators, with 
$s_1 \equiv \sigma_{-}$, $s_2 \equiv \sigma_{+}$, $q_{1} \equiv \sum\limits_{\textbf{k},\xi}g_{\textbf{k},\xi}^{\ast}b_{\textbf{k},\xi}^{\dagger}$, and $q_{2} \equiv \sum\limits_{\textbf{k},\xi}g_{\textbf{k},\xi}b_{\textbf{k},\xi}$. 
Then
\begin{widetext}
\begin{align}
H'_{SB}(t) &= \hbar\sum\limits_{m=1,2} e^{\frac{i}{\hbar}(H_{S} + H_{B})t}s_{m}q_m e^{-\frac{i}{\hbar}(H_{S} +H_{B})t} \nonumber\\
& = \hbar\sum\limits_{m=1,2} \left(e^{\frac{i}{\hbar}H_{S}t}s_{m}e^{-\frac{i}{\hbar}H_{S}t}\right)\left(e^{\frac{i}{\hbar}H_{B}t}q_m e^{-\frac{i}{\hbar}H_{B}t}\right) \nonumber\\
& = \hbar\sum\limits_{m=1,2} s_{m}(t) q_m(t)
\end{align}
where one can show using Baker-Campbell-Hausdorff formula that $s_1(t)=\sigma_-e^{-i\omega_0t}$, $s_2(t)=\sigma_+e^{i\omega_0t}$, 
$q_1(t)=\sum\limits_{{\bf k},\xi}g^\ast_{{\bf k},\xi}b^{\dagger}_{{\bf k},\xi}e^{i\omega_{\bf k}t}$, and 
$q_2(t)=\sum\limits_{{\bf k},\xi}g_{{\bf k},\xi}b_{{\bf k},\xi}e^{-i\omega_{\bf k}t}$. 
The master equation can be written as 
\begin{align}
\frac{d\rho^\prime(t)}{dt} = -\frac{i}{\hbar}[H^\prime_{\text{ext}}(t),\rho^\prime(0)]&-\frac{1}{\hbar^2}\int_{0}^{t} dt^{\prime} \Tr_B\left([H^\prime_{\text{ext}}(t),[H^\prime_{\text{ext}}(t^{\prime}),\rho^\prime(t^{\prime})\otimes \rho_{B}^{eq}]]\right)\nonumber\\
&- \frac{1}{\hbar^2}\int_{0}^{t} dt'
\left[ \left(s_1(t)s_2(t')\rho'(t') - s_2(t')\rho'(t')s_1(t)\right) \langle q_1(t)q_2(t') \rangle_B\right. \nonumber\\
&  \left. \qquad\qquad+ \left(s_2(t)s_1(t')\rho'(t') - s_1(t')\rho'(t')s_2(t)\right) \langle q_2(t)q_1(t') \rangle_B \right.\nonumber\\
&  \left. \qquad\qquad+ \left(\rho'(t')s_1(t')s_2(t) - s_2(t)\rho'(t')s_1(t')\right) \langle q_1(t')q_2(t) \rangle_B  \right.\nonumber\\
&  \left. \qquad\qquad+ \left(\rho'(t')s_2(t')s_1(t) - s_1(t)\rho'(t')s_2(t')\right) \langle q_2(t')q_1(t) \rangle_B\right], 
\end{align}
where $\langle q_m(t)q_n(t') \rangle_B = \text{Tr}_B\left[\rho_{B}^{eq} q_m(t)q_n(t') \right]$ represents the bath correlation function. 
For example, 
\begin{align}
\langle q_1(t)q_2(t')\rangle_B
&=\sum_{{\bf k},\xi}\sum_{{\bf k}',\xi'}
g_{{\bf k},\xi}^\ast g_{{\bf k}',\xi'} e^{i\omega_{\bf k}t}e^{-i\omega_{{\bf k}'}t'}
\text{Tr}_B(\rho_B^{eq}b_{{\bf k},\xi}^\dagger b_{{\bf k}',\xi'})\nonumber\\
&=\sum_{{\bf k}}
|g_{{\bf k},\xi}|^2 e^{i\omega_{\bf k}(t-t')}\bar{n}(\omega_{\bf k}). 
\end{align}
Likewise 
\begin{align}
\langle q_2(t)q_1(t')\rangle_B=\sum_{{\bf k}}
|g_{{\bf k}}|^2 e^{-i\omega_{\bf k}(t-t')}(\bar{n}(\omega_{\bf k})+1). 
\end{align}
By substituting $s_{1,2}$, $q_{1,2}$ back, and transforming the time $u=t-t'$, 
we obtain  
\begin{align}
\frac{d\rho^\prime(t)}{dt} =-\frac{i}{\hbar}[H^\prime_{\text{ext}}(t),\rho^\prime(0)]&-\frac{1}{\hbar^2}\int_{0}^{t} dt^{\prime} \Tr_B\left([H^\prime_{\text{ext}}(t),[H^\prime_{\text{ext}}(t^{\prime}),\rho^\prime(t^{\prime})\otimes \rho_{B}^{eq}]]\right)\nonumber\\
& - \sum_{{\bf k},\xi}|g_{{\bf k},\xi}|^2\int_{0}^{t} du
\left[ \left(\sigma_-\sigma_+\rho'(t-u) - \sigma_+\rho'(t-u)\sigma_-\right)e^{i(\omega_{\bf k}-\omega_0)u}\bar{n}(\omega_{\bf k})\right. \nonumber\\
&  \left. \qquad\qquad+ \left(\sigma_+\sigma_-\rho'(t-u) - \sigma_-\rho'(t-u)\sigma_+\right)e^{-i(\omega_{\bf k}-\omega_0)u} (\bar{n}(\omega_{\bf k})+1)\right.\nonumber\\
&  \left. \qquad\qquad+ \left(\rho'(t-u)\sigma_-\sigma_+ - \sigma_+\rho'(t-u)\sigma_-\right)e^{-i(\omega_{\bf k}-\omega_0)u}\bar{n}(\omega_{\bf k})\right.\nonumber\\
&  \left. \qquad\qquad+ \left(\rho'(t-u)\sigma_+\sigma_--\sigma_-\rho'(t-u)\sigma_+\right)e^{i(\omega_{\bf k}-\omega_0)u}(\bar{n}(\omega_{\bf k})+1) \right]. 
\end{align}
\end{widetext}
To proceed further, \emph{Markov approximation} is often considered with $\rho'(t-u)\rightarrow \rho'(t)$ and $\int_0^tdu \rightarrow \int_0^{\infty}du$ by assuming that the value of density matrix is determined without memory, and 
$\lim_{t\rightarrow \infty}\int_{-t}^tdu\sum_{{\bf k},\xi}|g_{{\bf k},\xi}|^2e^{\pm i(\omega_{\bf k}-\omega_0)u}\rightarrow \gamma\equiv 2\pi\int d^3kJ({\bf k})|g_{{\bf k},\xi}|^2\delta(\omega_{\bf k}-\omega_0)$ where $J({\bf k})$ is the spectral density, determining the decay rate $\gamma$ \cite{Carmichael}.

Transformed back to the Schr{\"o}dinger picture, the master equation takes the following form \cite{Carmichael}
\begin{align}
\frac{d\rho(t)}{dt}= - \frac{i}{\hbar}[H_S+H_\text{ext}(t), \rho(t)]+\mathcal{D}(\rho(t))
\end{align}
with the Lindblad dissipator 
\begin{align}
\mathcal{D}(\rho)&= \dfrac{\gamma}{2}(\bar{n}+1)\left( 2\sigma_-\rho\sigma_+ - \sigma_+\sigma_-\rho- \rho\sigma_+\sigma_- \right)\nonumber\\
&+\dfrac{\gamma}{2}\bar{n}\left( 2\sigma_+\rho\sigma_- - \sigma_-\sigma_+\rho - \rho\sigma_-\sigma_+ \right),
\label{eqn:Dissipator}
\end{align}
where $\bar{n}=\bar{n}(\omega_0)=(e^{\beta\hbar\omega_0}-1)^{-1}$ is the average number of thermal photons. 
\\

\setcounter{equation}{0}
\renewcommand{\theequation}{B\arabic{equation}}

\subsection{Transformation to the rotating frame}
The density matrix $\tilde{\rho}=\ket{\phi}\bra{\phi}$ in the rotating frame is transformed into the one in the stationary frame $\rho=\ket{\psi}\bra{\psi}$ via the following operation,
\begin{equation}
\underbrace{\ket{\psi}\bra{\psi}}_{=\rho} = e^{-i\omega t \sigma_{z}/2}\underbrace{\ket{\phi}\bra{\phi}}_{=\tilde{\rho}}e^{i\omega t \sigma_{z}/2}, 
\label{rot_trans1}
\end{equation}
which allows us to express the excited state population in the stationary frame into the one in the rotating frame as 
\begin{equation}
\tilde{\rho}_{ee} = e^{i\omega t \sigma_{z}/2}\underbrace{\ket{e}\bra{e}}_{=\rho_{ee}}e^{-i\omega t \sigma_{z}/2}. 
\label{rot_trans2}
\end{equation}
Employing the Baker-Campbell-Hausdorff formula, 
$e^{s\hat{A}}\hat{B}e^{-s\hat{A}} = \hat{B} + (s/1!)[\hat{A},\hat{B}] + (s^2/2!)[\hat{A},[\hat{A},\hat{B}]] + \cdots$, 
one can show that 
 $\tilde{\rho}_{ee}=\rho_{ee}$, $\tilde{\rho}_{gg}=\rho_{gg}$, $\tilde{\rho}_{eg}=\rho_{eg}e^{i\omega t}$, and $\tilde{\rho}_{ge}=\rho_{ge}e^{-i\omega t}$. 
 \\

\setcounter{equation}{0}
\renewcommand{\theequation}{C\arabic{equation}}

\begin{figure}[ht]
\includegraphics[width=0.75\columnwidth]{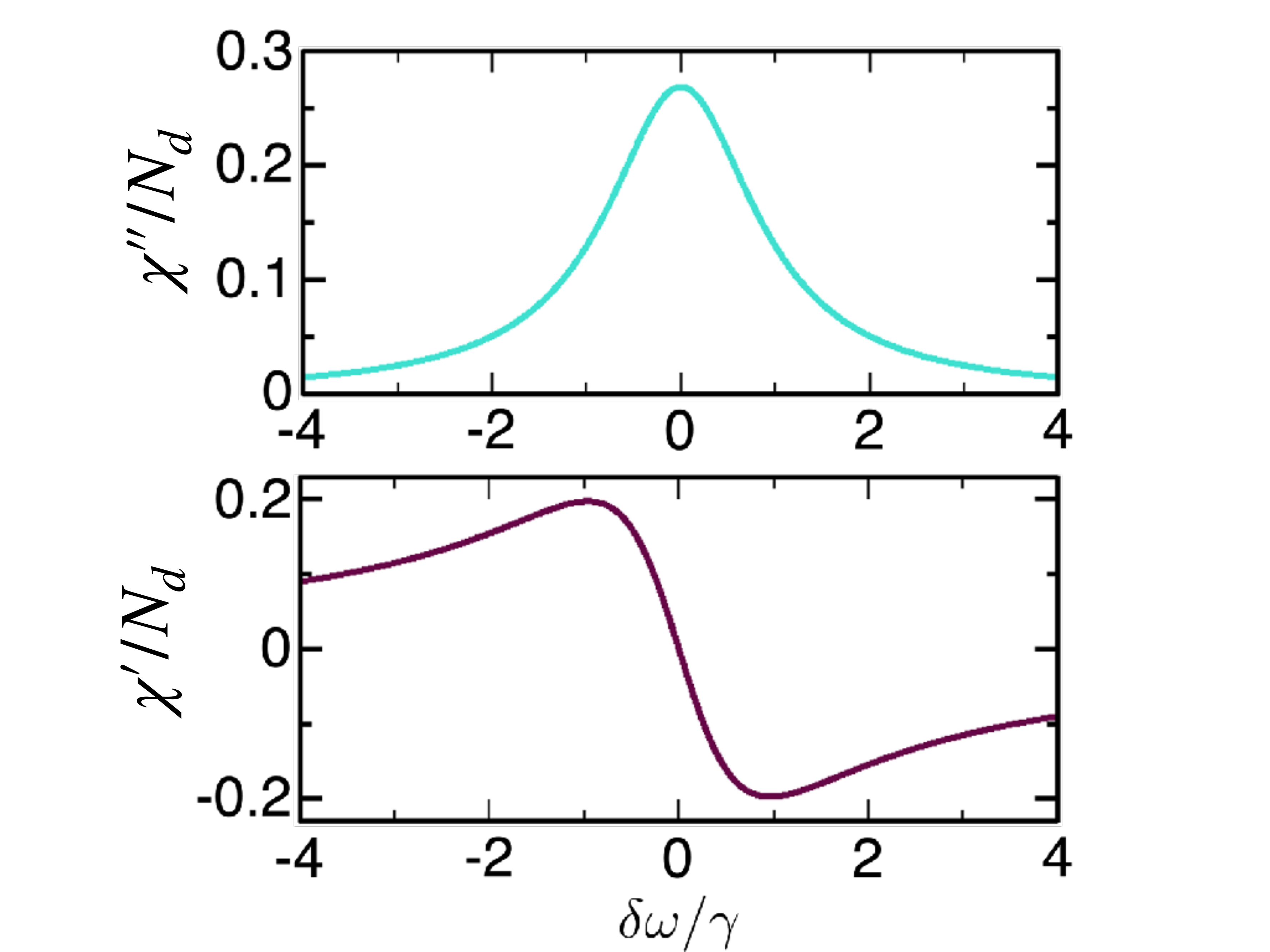}
\caption{The susceptibility of the system as a function of scaled detuning ($\delta\omega_\gamma$) calculated at $\Omega_\gamma = 0.5$ and $\mathcal{A}=\beta\hbar\omega_0 = 2$. (A) the absorption profile (i.e., $\chi''$) of the TLS. 
(B) the dispersion profile (i.e., $\chi'$) of the medium. 
}
\label{Absorption_dispersion}
\end{figure}

\subsection{Relation between the coherence and dielectric susceptibility}
Inside a dielectric medium, the electric polarization, defined as the total dipole moment per volume, i.e., $\vec{P} = N\langle \vec{d} \rangle$ with $N$ the number of molecules per volume, 
is induced by the incident radiation field, i.e., $\vec{P} =\chi\vec{E}$, where $\chi$ is the linear susceptibility of the medium \cite{scully_zubairy_1997}. 
With $\langle \vec{d} \rangle = \text{Tr}(\tilde{\rho}\vec{d}) = \tilde{\rho}_{eg}\vec{d}_{ge} + \tilde{\rho}_{ge}\vec{d}_{eg}= e^{i\omega t}(\rho_{eg}\vec{d}_{ge}+\rho_{ge}\vec{d}_{eg}e^{-2i\omega t})\approx \tilde{\rho}_{eg}\vec{d}_{ge}$, the susceptibility can be expressed as $\chi = N_d\tilde{\rho}_{eg}$ with $N_d = N|d_{ge}|/|\vec{\varepsilon}|$. In the light-matter interactions, the properties of the medium can be investigated by studying $\chi$, which is decomposed into the real ($\chi'$) and imaginary part ($\chi''$), i.e., $\chi=\chi'+i\chi''$.   
For the medium with $|\chi|\ll 1$, since the refractive index ($\underbar{n}$) is related with the dielectric constant as 
$\underbar{n}^2=\epsilon = 1 + 4\pi\chi$ (in Gaussian units), 
\begin{align}
\underbar{n} \approx 1 + 2\pi\chi'. 
\end{align}
The incident radiation field is absorbed to the medium is quantified by the attenuation coefficient $\alpha$, which is related to the wave vector as $k=\beta+i\alpha/2$, which leads to 
$\vec{E}\sim e^{ikz}\sim e^{i\beta z-\alpha z/2}$ and $|\vec{E}|^2\sim e^{-\alpha z}$. 
From $k=(\omega/c)\underbar{n}=(\omega/c)\sqrt{1+4\pi\chi'+4\pi i\chi''}\approx (\omega/c)(1+2\pi\chi'+2\pi i\chi'')$, it follows that  
\begin{align}
\alpha=4\pi\left(\frac{\omega}{c}\right)\chi''. 
\end{align}
This implies that the real part of the susceptibility (i.e., $\chi'= N_d \tilde{\rho}_R$) informs about the dispersion profile of the medium, while the imaginary part of the susceptibility (i.e., $\chi'' = N_d \tilde{\rho}_I$) describes the absorption profile of the medium. 

The degree of the absorption and dispersion varies with detuning of the field (Fig.~\ref{Absorption_dispersion}). 
For the perfect resonance (i.e., when $\delta\omega = 0$) the excitation is completely absorbed by the medium ($\chi''$ maximized in Fig.~\ref{Absorption_dispersion}) without any dispersion ($\chi'=0$ in Fig.~\ref{Absorption_dispersion}). 
With increasing $|\delta\omega|$, 
the absorption (dispersion) in the medium decreases (increases). 

\setcounter{equation}{0}
\renewcommand{\theequation}{D\arabic{equation}}

\subsection{Method of generating function}
The dynamical equations (Eq.~\ref{dynamical_equation2}) can be cast into the matrix form,  
\begin{align}
\partial_\tau\tilde{\rho}(\tau)=\mathcal{L}\tilde{\rho}(\tau). 
\label{eqn:matrixform}
\end{align}
where $\tilde{\rho}=(\tilde{\rho}_{ee},\tilde{\rho}_{eg},\tilde{\rho}_{ge},\tilde{\rho}_{gg})^T$ is a vector in Liouville space, and $\mathcal{L}$ denotes the $4\times 4$ Liouvillian super-operator. 
{\color{black}We aim to count} the net number of photon transfer to the surrounding bath for time interval $\tau$ ($n(\tau)$) 
and calculate its mean current and current fluctuations at steady state.  
{\color{black}Since the light-induced absorption and emission is a cyclic process for the system, we assume that the system is periodic in state space and that the superoperator $\mathcal{L}$ is periodic with its period by $L(=4)$, satisfying $\mathcal{L}_{ij}=\mathcal{L}_{i+L,j+L}$ for all $i$, $j=0$, $1$, $2$, $3$ representing the quantum state $ee$, $eg$, $ge$, $gg$, respectively, and  
introduce a generalized coordinate $\mu\in\mathbb{Z}$}, which is related with {\color{black}the numeric index $i=0$, $1$, $2$, $3$ } as $\mu= i\text { }(\text{mod}\text{ }L)\text{ with }L=4$. 
This allows us to count the cycles of absorptions followed by emissions via $\tilde{\rho}_i(\mu,\tau)\equiv \tilde{\rho}(\mu,\tau)\delta_{\mu,i}^L$ with the generalized Kronecker delta, $\delta_{\mu,i}^L=1$ if $\mu=i$ (mod $L$) and $\delta_{\mu,i}^L=0$ otherwise ~\cite{Koza1999JPA,bruderer_inverse_2014}. 
Here, $\tilde{\rho}_i(\mu,\tau)$ denotes the probability of finding a system at a given quantum state $i\in\{ee,eg,ge,gg\}$ at site $\mu$ at time $\tau$. 

We consider a generating function for each element $\tilde{\rho}_i$ of the vector $\tilde{\rho}$,  
\begin{align}
\mathcal{G}_i(z,\tau)=\sum_{\mu=-\infty}^{\infty}e^{z n_\mu}
\tilde{\rho} _i(\mu,\tau).   
\label{eqn:Generating}
\end{align}
Multiplying the factor $e^{zn_\mu}$ to both sides of a generalized version of Eq.~\ref{eqn:matrixform}, namely $\partial_\tau\tilde{\rho}_i(\mu,\tau)=\sum_{j}\mathcal{L}_{ij}\tilde{\rho}_j(\mu,\tau)$, and summing over $\mu$, 
one gets  
\begin{align}
\partial_\tau\mathcal{G}_i(z,\tau)=\sum_j[\Gamma(z)]_{ij}\mathcal{G}_j(z,\tau)
\label{eqn:ODE_generating}
\end{align}
where 
\begin{widetext}
\begin{align}
\Gamma(z) &\equiv  \begin{bmatrix} -(\bar{n} + 1) & -i\Omega_\gamma & i\Omega_\gamma & \bar{n}e^{-z} \\
-i\Omega_\gamma & i\delta\omega_\gamma - (2\bar{n} + 1)/2 & 0 & i\Omega_\gamma \\
i\Omega_\gamma & 0 & -i\delta\omega_\gamma - (2\bar{n} + 1)/2 & -i\Omega_\gamma \\
(\bar{n} + 1)e^{z} & i\Omega_\gamma & -i\Omega_\gamma & -\bar{n} \end{bmatrix}.
\end{align}
\end{widetext}
In obtaining the matrix $\Gamma(z)$ where the factor $e^{z n_{\mu}}$ is multiplied to each element, we have set $n_{\mu}=-1$ for the matrix element $[\Gamma(z)]_{ee,gg}$ corresponding to absorption, $n_{\mu}=1$ for $[\Gamma(z)]_{gg,ee}$ corresponding to emission, and $n_{\mu}=0$ for the rest.   
The formal solution of Eq.~\ref{eqn:ODE_generating} can be written as 
\begin{align}
\mathcal{G}_i(z,\tau)=\sum_{\alpha}T_{\alpha i}(z)e^{\lambda_\alpha(z)\tau}
\end{align}
where $\lambda_{\alpha}(z)$ and $T_{\alpha i}(z)$ are the $\alpha$-th eigenvalue and the corresponding eigenstate of $\Gamma(z)$, 
satisfying $\sum_j\left[\Gamma(z)\right]_{ij}T_{\alpha j}=\lambda_\alpha(z)T_{\alpha i}(z)$ with $\lambda_0(z)>\lambda_1(z)>\lambda_2(z)>\lambda_3(z)$ ($\alpha=0,1,2,3$).   
At steady state, 
$\mathcal{G}_i(z,\tau)$ is dominated by the term with the largest eigenvalue 
\begin{align}
\lim_{\tau\rightarrow\infty}\mathcal{G}_i(z,\tau)\sim T_{0i}(z,\tau)e^{\lambda_0(z)\tau}. 
\end{align}
\begin{widetext}
The eigenvalues $\lambda_\alpha(z)$ of $\Gamma(z)$ are obtained from the characteristic polynomial 
\begin{align}
 \det | \lambda(z)\mathcal{I}-\Gamma(z)|= \sum_{n=0}^4a_n(z)\lambda^n(z)=0,
\label{pol}
\end{align}
where 
\begin{align}
a_4 &=1\nonumber\\
a_3 &=2(2\bar{n} + 1)\nonumber\\
a_2 &= \frac{1}{4}\left[ 5(2\bar{n} + 1)^2 + 4\delta\omega_\gamma^2 + 16\Omega_\gamma^2 \right]  \nonumber\\
a_1(z) &= \frac{1}{4}\left[ (2\bar{n} + 1)^3 + 4\delta\omega_\gamma^2(2\bar{n} + 1) + 16\Omega_\gamma^2(2\bar{n} + 1) - 8\Omega_\gamma^2\left( e^{z} + \bar{n}e^{-z} + \bar{n}e^{z}\right)\right]  \nonumber\\
a_0(z) &= -\Omega_\gamma^2\left[ e^{z} + \bar{n}e^{-z} + 2\bar{n}^2e^{-z} + 3\bar{n}e^{z} + 2\bar{n}^2e^{z} - (2\bar{n} + 1)^2 \right].\nonumber
\end{align}
\end{widetext}

Now, summing $\mathcal{G}_i(z,\tau)$ over the index $i\in\{ee,eg,ge,gg\}$, we obtain the following expression that can be used as the moment generating function: 
\begin{align}
\mathcal{G}(z,\tau)\equiv \sum_i\mathcal{G}_i(z,\tau)=\sum_{i}\sum_{\mu=-\infty}^{\infty}e^{zn_\mu}\tilde{\rho}_i(\mu,\tau)
\label{eqn:moment_generating}
\end{align}
which allows us to calculate the first and second cumulant of $n(\tau)$ as follows. 
\begin{align}
\langle n(\tau)\rangle=\frac{\sum_i\sum_\mu n_\mu\tilde{\rho}_i(\mu,\tau)}{\sum_i\sum_\mu \tilde{\rho}_i(\mu,\tau)}=\partial_z\log{\mathcal{G}(z,\tau)}|_{z=0}
\label{eqn:X}
\end{align}
and 
\begin{align}
\langle n(\tau)^2\rangle-\langle n(\tau)\rangle^2=\partial_z^2\log{\mathcal{G}(z,\tau)}|_{z=0}. 
\label{eqn:dX2}
\end{align}

Finally, from Eqs.~\ref{eqn:X} and \ref{eqn:dX2}, and the asymptotic expression of $\mathcal{G}(z,\tau)$, 
\begin{align}
\lim_{\tau\rightarrow\infty}\mathcal{G}(z,\tau)\sim h_0(z,\tau)e^{\lambda_0(z)\tau}. 
\end{align}
with $h_0(z,\tau)\equiv \sum_iT_{0i}(z,\tau)$, 
we can obtain $\langle j\rangle$ and $\text{Var}[j]$, 
\begin{align}
\langle j\rangle \equiv \lim_{\tau\rightarrow\infty}\frac{\langle n(\tau)\rangle}{\tau} =\lambda_0'(0)
\end{align}
and 
\begin{align}
\text{Var}[j] \equiv\lim_{\tau\rightarrow\infty}
\frac{\langle n(\tau)^2\rangle-\langle n(\tau)\rangle^2}{\tau} =\lambda_0''(0). 
\end{align}
These key values of $\lambda'_0(0)$ and $\lambda''_0(0)$ can be evaluated in terms of the coefficients of the characteristic polynomial (Eq.~\ref{pol}) differentiated with respect to $z$ at $z=0$ as follows: 
\begin{align}
a_0'(0)+a_1(0)\lambda_0'(0)=0
\end{align}
and 
\begin{align}
a_0''(0)+2a_1'(0)\lambda_0'(0)+a_1(0)\lambda_0''(0)+2a_2(0)(\lambda_0'(0))^2=0.
\end{align}
Therefore, 
\begin{align}
\langle j\rangle=\lambda_0'(0)=-\frac{a_0'(0)}{a_1(0)}
\end{align}
and 
\begin{align}
&\text{Var}[j]=\lambda_0''(0)\nonumber\\
&=-\frac{1}{a_1(0)}\left(a_0''(0)+2a_1'(0)\lambda_0'(0)+2a_2(0)(\lambda_0'(0))^2\right). 
\end{align}

\bibliography{Paper_TUR,mybib1}

\end{document}